\documentclass[aps,prd,showpacs,showkeys,12pt]{revtex4}
\usepackage{amsmath,amssymb} 
\usepackage[dvips]{graphicx} 
\usepackage{graphics}
\usepackage{graphicx}
\usepackage{epsfig}
\usepackage[english]{babel}
\newcommand\nn{\nonumber}
\newcommand\ba{\begin{eqnarray}}
\newcommand\ea{\end{eqnarray}}
\newcommand\ee{\end{equation}}
\newcommand\be{\begin{equation}}

\begin{document}

\title{Production of $\pi^0\rho^0$ pair in electron-positron annihilation in the
Nambu-Jona-Lasinio model}

\author{A.~I.~Ahmadov}
\email{ahmadov@theor.jinr.ru}
\affiliation{Joint Institute for Nuclear Research, Dubna, Russia}
\affiliation{Institute of Physics, Azerbaijan National Academy of Sciences, Baku, Azerbaijan}

\author{E.~A.~Kuraev}
\email{kuraev@theor.jinr.ru}
\affiliation{Joint Institute for Nuclear Research, Dubna, Russia}

\author{M.~K.~Volkov}
\email{volkov@theor.jinr.ru}
\affiliation{Joint Institute for Nuclear Research, Dubna, Russia}

\begin{abstract}
The process $e^+e^- \to \pi^0\rho$ is described in the framework of the expanded NJL model in the
energy region from 0.9 GeV to 1.5 GeV. The contribution of intermediate state with vector mesons
$\omega(782), \,\,\phi(1020)$, and $\omega'(1420)$, where $\omega'$ is the first radial excitation
of $\omega$ - meson was taken into account.
Results obtained are in satisfactory agreement with experimental data.
\end{abstract}

\date{\today}



\maketitle


\section{Introduction}
\label{Introduction}
This work is devoted to the description of the process  $e^+e^- \to \pi^0\rho$ recently measured in
experiments of 2006 and 2011 years ~\cite{akhmetshin1,akhmetshin2}.
It is also concerned with a theoretical study of the process
$e^+e^- \to \pi^0\rho^0$ within the $\omega-, \omega'-$ and $\phi-$ mesons energy range.
This process was recently measured at the CMD-2 detector at the VEPP-2M $e^+e^-$ collider ~\cite{akhmetshin1,akhmetshin2,achasov}.

The cross section of hadron production in the $e^+e^-$ annihilation in the energy region $\sqrt {s} < 1.03 \,\,GeV$
can be described within the vector meson dominance model (VDM) framework and is determined by the transitions of
light vector mesons ($\omega, \omega', \phi$) to the final states.

It is one of the series of works ~\cite{volkov1,volkov2}, where process
$e^+e^- \to \pi^0 \omega,\,\,\,\pi^0\gamma$ was described in the framework of the expanded NJL model
~\cite{volkov8,volkov9}.
The results obtained were found to be in satisfactory agreement with known experimental data ~\cite{akhmetshin1,akhmetshin2}.
The main formalism including the $SU(2) \times SU(2)$ chiral NJL model coincides with
one of paper ~\cite{volkov1,volkov2}.

The standard NJL Lagrangian which describes interactions of photons, pions and vector $\rho$ and $\omega$
mesons with quarks, see Refs.~\cite{volkov7,volkov5}.

\section{Amplitude of the process $e^+e^- \to \omega, \omega', \phi \to \pi^0\rho^0$}
\label{Amplitude}

The amplitude can be written down in the form:

\ba
T=\bar {e}\gamma_{\mu}e \cdot \varepsilon_{\mu\lambda\alpha\beta} \frac{p_{\pi}^{\alpha}p_{\rho}^{\beta}}{m \cdot s}\cdot
\{B_{\gamma+\omega}+B_{\phi}+B_{\omega'}\}\varepsilon_{\lambda}(\rho)
\ea
where $s=(p_1(e^+)+p_2(e^-))^2$.

The quantity $B_{\gamma+\omega}$ to the contribution of the amplitude from the process with intermediate photons and $\omega$ - mesons:
\ba
B_{\gamma+\omega}=\frac{M_{\omega}^2 +i M_{\omega}\Gamma_{\omega}}{M_{\omega}^2 -s +iM_{\omega}\Gamma_{\omega}} \cdot
\frac{1}{g_{\rho_1}}\cdot V_{\rho\pi^0\gamma}(s).
\ea

The quantity $B_{\phi}$ corresponds to the contribution with $\phi$ - meson in the intermediate state ~\cite{volkov9}:
\ba
B_{\phi}=\frac{s\sqrt{2} \cdot \sin\theta_{\omega\phi}}{s-M_{\phi}^2 +iM_{\phi}\Gamma_{\phi}} \cdot
\frac{1}{g_{\rho_1}}\cdot V_{\rho\pi^0\gamma}(s),
\ea
$\sin\theta_{\omega\phi} = -0.0523.$  \\
The quantity $B_{\omega'}$ to the contribution from the intermediate radial excitation of the  $\omega$ - meson state:
$\omega' \to \pi^0\rho$ it is taken from paper ~\cite{volkov3}
\ba
B_{\omega'}=\frac{s}{s-M_{\omega'}^2 \Gamma_{\omega'}} \biggl(-\frac{\cos(\beta+\beta_0)}{\sin(2\beta_0)}-
\Gamma \frac{\cos(\beta-\beta_0)}{\sin(2\beta_0)}\biggr)\cdot \frac{1}{g_{\rho_1}}\cdot V_{\rho'\pi^0\gamma}(s),
\ea
where \\
$\Gamma  \approx \frac{1}{2}$ will be specified below (see (8)) and
\ba
V_{\rho\pi^0\gamma}(s) = g_{\pi_1} \biggl(\frac{\sin(\beta+\beta_0)g_{\rho_1}I_0^{(3)}}{\sin(2\beta_0)}+
\frac{\sin(\beta-\beta_0)g_{\rho_2}I_1^{(3)}}{\sin(2\beta_0)}\biggr)\frac{1}{g_{\rho_1}} V_{\rho'\pi^0\gamma}, \nn \\
V_{\rho'\pi^0\gamma}(s) = -g_{\pi_1} \biggl(\frac{\cos(\beta+\beta_0)g_{\rho_1}I_0^{(3)}}{\sin(2\beta_0)}+
\frac{\cos(\beta-\beta_0)g_{\rho_2}I_1^{(3)}}{\sin(2\beta_0)}\biggr),
\ea
\ba
I_n^{(3)} =-\int\frac{d^4k \cdot m^2 f^n(k^{\bot^2})\Theta(\Lambda^2 -|k^{\bot^2}|)}{i\pi^2(k^2-m^2+i0)} \cdot \nn \\
\frac{1}{((k+p_{\rho})^2-m^2+i0)\cdot ((k+p_{\pi})^2-m^2+i0)}.
\ea
The radially-excited states were introduced in the NJL model with the help of the form factor in the quark-meson
interaction:
\ba
f(k^{\bot^2}) = (1-d|k^{\bot^2}|)\Theta(\Lambda^2-|k^{\bot^2}|), \nn \\
k^{\bot}=k-\frac{(kp)p}{p^2}, \,\,\,\, d=1.78 \,\,\,GeV^{-2},
\ea
where $k$ and $p$ are the quark and meson momenta, respectively. The cut-off  parameter $\Lambda$=1.03 \,\,GeV
is taken ~\cite{volkov4}. The coupling constants $g_{\pi_1}=g_{\pi}$ and $g_{\rho_1}=g_{\rho}$ are the same as in the standard NJL version. The constants $g_{\pi_2}=3.20, \,\,\,g_{\rho_2}=9.87$ are the mixing angles $\beta_0 =61.53^{\circ}$, and $\beta =76.78^{\circ}$ were defined in ~\cite{volkov3}. The standard value of the $\phi-\omega$ mixing angle $\theta_{\omega\phi} \approx -3^{\circ}$ is used ~\cite{volkov5}.
So for the numerical calculations we use the values from the Particle data Group ~\cite{pdg}:
$\Gamma_{\omega}=8.49 \,\,\,MeV, \,\,\,\Gamma_{\omega'}=215 \,\,\,MeV, \,\,\,M_{\omega}=782 \,\,\,MeV, \,\,\,M_{\rho}=775 \,\,\,MeV, \,\,\,
M_{\omega'}=1420 \,\,\,MeV, \,\,\, M_{\phi}=1020 \,\,\,MeV, \,\,\, \Gamma_{\phi}=4.26 \,\,\,MeV, \,\,\, M_{\pi}=139 \,\,\,MeV$.
The $\gamma-\omega$ transition differs from the above just by factor $1/3$ compared with $\gamma-\rho$. In the amplitudes with excited mesons we have to take into account the $\gamma-\rho_2$ and $\gamma-\omega_2$ transitions ($\gamma-\omega_1(\rho_1)$ ones are the same as in the standard $\gamma-\omega(\rho)$ cases) can be expressed
via the $\gamma-\omega(\rho)$ transition with the additional factor ~\cite{volkov9, volkov3}
\ba
\Gamma=\frac{I_2^f}{\sqrt{I_2I_2^{f^2}}} \approx 0.47.
\ea
\section{Total Cross Section}
\label{TotalCrossSection}
In (7) $m$ is the constituent quark mass ($m_u=m_d =280 \,\,\,MeV$).
For calculation of the total cross section of the process we use:
\ba
\sigma(s) = \frac{3\alpha^2 g_{\rho}^2}{32 \pi^3 s^3 f_{\pi}^2} \lambda^{3/2}(s, M_{\rho}^2, M_{\pi}^2)
\cdot
|B_{\gamma+\omega} +B_{\phi} +B_{\omega'}|^2,
\ea
where $f_{\pi} = 93 \,\,\,MeV$ is the pion decay constant and $\lambda(s, M_{\rho}^2, M_{\pi}^2)=
(s-M_{\pi}^2 -M_{\pi}^2)^2-4 M_{\rho}^2 M_{\pi}^2$, $g_{\rho}$ is the vector meson coupling constant
$g_{\rho} \approx $ 6.14 corresponding to the standard relation $g_{\rho}^2 \approx$ 3.

The total cross section in the region $0.9 \,\,\,GeV < \sqrt {s} < 2 \,\,\,GeV$ is presented in Fig.1.

\begin{figure}
\includegraphics[width=0.9\textwidth]{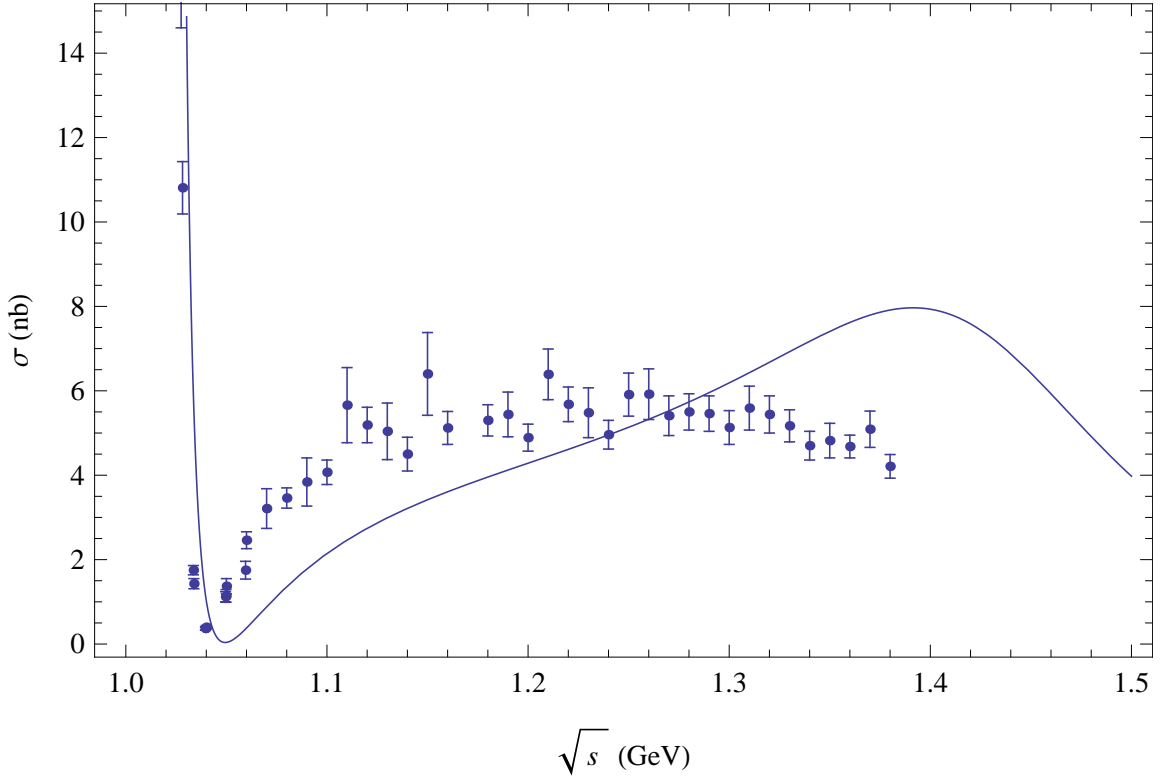}
\caption{Total cross section as a function $\sqrt {s}, \,\,\, 1.02 < \sqrt {s} <2 \,\,\,GeV$ of the $e^+e^- \to \pi^0\rho^0$
process in the NJL model. Points are experimental data ~\cite{achasov}.
\label{Fig1}}
\end{figure}
In Table~\ref{Table1} the behavior of the cross section in the region $m_{\pi}+m_{\rho} = \sqrt {s_{th}} < \sqrt {s} =1.1 \,\,\, GeV$
is presented. In this region the cross section has a resonance character. \\
In conclusion, we would like to note the distinction of between the $\pi^0\rho^0$ and $\pi^0 \omega$ process,
where the $\phi$ - resonance is not seen in the $\pi^0 \omega$ process. A similar situation takes place in the
process $e^+e^- \to \pi^0\gamma$ which was supported by experimental data.

As a by product of our analysis we obtain the partial decay of the process $\phi \to \rho^0\pi^0$,
$\Gamma_{\phi \to \rho^0\pi^0} \approx 0.5 MeV$ which is in good agreement with PDG data ~\cite{pdg}.

\newpage

\begin{table}
\begin{tabular}{|c|c|c|c|c|c|c|c|c|c|c|c|c|c|c|}
\hline
$\sqrt {s} (GeV)$ & 0.915 & 0.916 & 0.918 & 0.922 & 0.926 & 0.932 & 0.944 & 0.95 & 0.956 & 0.962 & 0.972 & 0.98 & 1 & 1.01   \\
\hline
$\sigma (nb)$ & 0 & 0.022 & 0.11 & 0.38 & 0.71 & 1.3 & 2.2  & 3.4 & 4.27 & 5.2 & 7.1 & 9.11 & 22 & 58.6 \\
\hline
\end{tabular}
\vspace*{0.2cm}
\begin{tabular}{|c|c|c|c|c|c|c|c|c|c|c|c|c|c|c|}
\hline
$\sqrt {s} (GeV)$ & 1.02 & 1.026 & 1.03 & 1.04 & 1.048 & 1.05 & 1.052 & 1.054 & 1.056 & 1.06 & 1.07 & 1.08 & 1.09 & 1.1 \\
\hline
$\sigma (nb)$ & 796 & 58.2 & 15.8 & 1.03 & 0.05 & 0.04 & 0.07 & 0.13 & 0.2 & 0.38 & 0.88 & 1.36 & 1.78 & 2.14 \\
\hline
\end{tabular}
\caption{The magnitude of the total cross section in the resonance region $0.915 < \sqrt {s} < 1.1 \,\,\,GeV$}
\label{Table1}
\end{table}
\section{Conclusions}
\label{Conclusions}
The cross section of the process $e^+e^- \to \pi^0\rho^0$ was measured in the Spherical Neutral Detector
(SND) experiment at the VEPP-2M collider in the energy region $\sqrt {s} =980 - 1380 \,\,MeV$
~\cite{akhmetshin1,akhmetshin2,achasov}.

Our calculations for the process $e^+e^- \to \pi^0\rho^0$ showed the presence of two regions of enhancement
of the cross section in the energy range below 1.020 \,\,\,GeV and 1.4 \,\,\,GeV. The first one appears
in the region of the $\phi$ meson mass and looks like a very high narrow peak. The second one is a smooth peak,
it lies in the region of $\omega'$ meson mass.

Notwithstanding, the process $e^+e^- \to \pi^0\rho^0$ is similar to the process $e^+e^- \to \omega\pi^0$,
but in our result in $\phi$ meson mass region we have a very narrow peak which will be agreement with experiment.
\acknowledgements

We would like to acknowledge the support of RFBR, grant no. 10-02-01295a.
This work was also supported by the Heisenberg--Landau program, grant HLP-2010-06 and the JINR--Belorus--2010 grant.
%


\end{document}